\documentclass[floats,floatfix,showpacs,amssymb,prd,twocolumn,superscriptaddress,nofootinbib,nolongbibliography,reprint]{revtex4-1}

\usepackage{amssymb,amsmath,verbatim,mathtools,needspace,enumitem,etoolbox,graphicx,physics,microtype,afterpage,xspace,tabularx,lmodern,multirow}
\usepackage{gensymb}
\usepackage[normalem]{ulem}
\usepackage[dvipsnames, usenames]{xcolor}
\usepackage{xr-hyper}
\definecolor{linkcolor}{rgb}{0.0,0.3,0.5}
\usepackage[unicode, colorlinks=true, linkcolor=linkcolor, citecolor=linkcolor, filecolor=linkcolor, urlcolor=linkcolor, linktocpage, breaklinks]{hyperref}
\usepackage[all]{hypcap}
\usepackage[T1]{fontenc}
\usepackage[utf8]{inputenc}
\usepackage[usenames,dvipsnames]{xcolor}
\hypersetup{colorlinks=true,citecolor=romared,linkcolor=romared,urlcolor=romared}

\definecolor{romared}{RGB}{142,0,28}

\newcommand{\be}{\begin{equation}}
\newcommand{\ee}{\end{equation}}

\def\be{\begin{equation}}
\def\ee{\end{equation}}
\newcommand{\beq}{\begin{eqnarray}}
\newcommand{\eeq}{\end{eqnarray}}

\usepackage{makecell}
\usepackage{soul}

\newcolumntype{Y}{>{\centering\arraybackslash}X}

\renewcommand{\vec}[1]{\boldsymbol{#1}}

\usepackage{orcidlink}

\newcommand{\orcid}[1]{\href{https://orcid.org/#1}{\includegraphics[width=10pt]{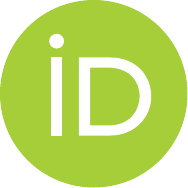}}}

\makeatletter
\newcommand*{\addFileDependency}[1]{
  \typeout{(#1)}
  \@addtofilelist{#1}
  \IfFileExists{#1}{}{\typeout{No file #1.}}
}
\makeatother

\newcommand*{\myexternaldocument}[1]{%
    \externaldocument{#1}%
    \addFileDependency{#1.tex}%
    \addFileDependency{#1.aux}%
}

\myexternaldocument{Supp}

\begin{document}

\title{Compact Binary Foreground Subtraction \\ in Next-Generation Ground-Based Observatories}

\author{Bei Zhou \orcid{0000-0003-1600-8835}}
\affiliation{William H. Miller III Department of Physics and Astronomy, Johns Hopkins University, Baltimore, Maryland 21218, USA}
\author{Luca Reali \orcid{0000-0002-8143-6767}}
\affiliation{William H. Miller III Department of Physics and Astronomy, Johns Hopkins University, Baltimore, Maryland 21218, USA}
\author{Emanuele Berti \orcid{0000-0003-0751-5130}}
\affiliation{William H. Miller III Department of Physics and Astronomy, Johns Hopkins University, Baltimore, Maryland 21218, USA}
\author{Mesut \c{C}al{\i}\c{s}kan \orcid{0000-0002-4906-2670}}
\affiliation{William H. Miller III Department of Physics and Astronomy, Johns Hopkins University, Baltimore, Maryland 21218, USA}
\author{Cyril Creque-Sarbinowski \orcid{0000-0002-6197-5421}}
\affiliation{William H. Miller III Department of Physics and Astronomy, Johns Hopkins University, Baltimore, Maryland 21218, USA}
\author{Marc Kamionkowski \orcid{0000-0001-7018-2055}}
\affiliation{William H. Miller III Department of Physics and Astronomy, Johns Hopkins University, Baltimore, Maryland 21218, USA}
\author{B. S. Sathyaprakash \orcid{0000-0003-3845-7586}}
\affiliation{Institute for Gravitation and the Cosmos, Department of Physics, Penn State University, University Park, Pennsylvania 16802, USA}
\affiliation{Department of Astronomy and Astrophysics, Penn State University, University Park, Pennsylvania 16802, USA}
\affiliation{School of Physics and Astronomy, Cardiff University, Cardiff, CF24 3AA, United Kingdom}

\date{\today}

\begin{abstract}
The stochastic gravitational-wave backgrounds (SGWBs) for current detectors are dominated by binary black-hole (BBH) and binary neutron-star (BNS) coalescences.  The sensitivity of current networks of gravitational-wave (GW) detectors allows only a small fraction of BBHs and BNSs to be resolved and subtracted, but previous work indicated that the situation should significantly improve with next-generation (XG) observatories. We revisit these conclusions by taking into account waveform-modeling uncertainties, updated astrophysical models, and (crucially) the full set of parameters that must be estimated to remove the resolved sources. Compared to previous studies, we find that the residual background from BBHs and BNSs is large even with XG detector networks. New data analysis methods will thus be required to observe the SGWB from cosmic supernovae or contributions from early-Universe phenomena like cosmic strings, stiff post-inflation fluids, or axion inflation.
\end{abstract}

\maketitle
\noindent {\bf \em Introduction.} 
Since the first detection in 2015, the LIGO/Virgo/KAGRA (LVK) network~\cite{Harry:2010zz, LIGOScientific:2014pky,VIRGO:2014yos,Aso:2013eba} has observed about 100 gravitational-wave (GW) events produced by the coalescence of black holes and neutron stars~\cite{LIGOScientific:2021djp,LIGOScientific:2021psn,Nitz:2021zwj,Olsen:2022pin}.
In addition to these loud and individually resolved GW events, a plethora of signals from multiple kinds of sources remain too weak to be detected. Their incoherent superposition produces stochastic GW backgrounds (SGWBs)~\cite{Allen:1997ad,Sathyaprakash:2009xs,Caprini:2018mtu,Christensen:2018iqi,Renzini:2022alw}. These SGWBs can be either of astrophysical origin (e.g., unresolved compact binaries~\cite{Wu:2011ac,Marassi:2011si,Zhu:2011bd,Rosado:2011kv,Zhu:2012xw, Dominik:2014yma} and supernova explosions~\cite{Ferrari:1998ut, Buonanno:2004tp, Crocker:2015taa, Crocker:2017agi, Finkel:2021zgf}) or of cosmological origin (produced e.g. by standard inflation~\cite{Grishchuk:1974ny,Starobinsky:1979ty,Grishchuk:1993te},  axion inflation~\cite{Barnaby:2011qe}, cosmic strings~\cite{Damour:2004kw, Siemens:2006yp, Olmez:2010bi, Regimbau:2011bm}, etcetera).

The current second-generation detector network did not detect any SGWB, placing upper bounds on the amplitude of the energy-density spectrum in various frequency bands~\cite{KAGRA:2021kbb}. Future SGWB detections may have a great scientific payoff. Astrophysical SGWBs potentially contain key information about the mass and redshift distributions and other properties of their sources~\cite{LIGOScientific:2020kqk, LIGOScientific:2021psn, Bavera:2021wmw}, while observing cosmological SGWBs would open a unique window to the earliest moments of the Universe and to the physical laws that apply at the highest energies~\cite{Grishchuk:1974ny,Starobinsky:1979ty,Grishchuk:1993te, Barnaby:2011qe, Damour:2004kw, Siemens:2006yp, Olmez:2010bi, Regimbau:2011bm}.

The predicted energy densities of different SGWBs vary by many orders of magnitude. The SGWB for current detectors is most likely dominated by a compact binary coalescence (CBC) ``foreground'', because the network's sensitivity allows only a small fraction of CBCs to be resolved and subtracted~\cite{Sharma:2020btq,Biscoveanu:2020gds}. Therefore, detection of the subdominant astrophysical and cosmological SGWBs cannot be accomplished. The situation could be significantly improved in next-generation (XG) observatories, including Cosmic Explorer (CE)~\cite{Reitze:2019iox} and the Einstein Telescope (ET)~\cite{Punturo:2010zz}, because their increased sensitivity should allow us to detect hundreds of thousands of BBHs and BNSs per year and better measure their parameters~\cite{Borhanian:2022czq,Ronchini:2022gwk,Iacovelli:2022bbs}.

Besides the contribution from the superposition of unresolved CBC GW signals ($\Omega_{\rm unres}$), there is also an unavoidable contribution coming from the fact that parameter estimation errors lead to imperfect subtraction of the resolvable events ($\Omega_{\rm err}$). The sum of these two residual backgrounds determines how well other subdominant SGWBs may be detected, and it must be minimized. 
Previous work argued that the increased sensitivity of XG networks would allow us to subtract CBC foregrounds much more precisely, and to search for subdominant SGWBs~\cite{Regimbau:2016ike, Sachdev:2020bkk}.

In this work, we update those studies in three important ways: (i) we consider populations of BNSs and BBHs with local merger rates consistent with the latest LVK catalog, and we take into account astrophysical uncertainties on these rates~\cite{LIGOScientific:2021djp,LIGOScientific:2021psn}; (ii) we carry out a preliminary exploration of waveform modeling systematics in the subtraction of CBC foregrounds; and
(iii) most importantly, we expand the range of binary parameters assumed to be determined from each GW signal. The details of our astrophysical population models and GW parameter estimation techniques are presented in a companion paper~\cite{ZhouPRD}. Previous work focused on the effect of the three dominant phase parameters (the redshifted detector-frame chirp mass $\mathcal{M}_z$, coalescence phase $\phi_c$, and time of coalescence $t_c$), while we consider the larger (9-dimensional) parameter space characterizing nonspinning binaries. We find that correlations and degeneracies between different parameters, and in particular the uncertainty in determining the amplitude of the individual signals, result in a much larger value for $\Omega_{\rm err}$ than previous estimates. We also find that the BBH residual background dominates the BNS background at low frequencies. New data analysis methods are required to observe the SGWB from cosmic supernovae or early-Universe backgrounds produced by cosmic strings, stiff post-inflation fluids, or axion inflation.

Throughout this work, $G$ is the gravitational constant, $c$ the speed of light, $H_0$ the Hubble constant, and we use the $\Lambda$CDM cosmological model with cosmological parameters taken from Planck 2018~\cite{Planck:2018vyg}.

\noindent {\bf \em Astrophysical population model.} 
Our sampled BBH and BNS event catalogs use state-of-the-art phenomenological models of the LVK population. 
The source-frame component masses $m_1$, $m_2$ of each BNS are sampled using the preferred model from Ref.~\cite{Farrow:2019xnc}, where the primary mass follows a double Gaussian distribution and the secondary mass is sampled uniformly. For BBH masses, we use the \texttt{POWER+PEAK} phenomenological model from the latest LVK population paper~\cite{LIGOScientific:2021psn}.

We consider the same BNS and BBH redshift distributions as in Refs.~\cite{Regimbau:2016ike,Sachdev:2020bkk}. We assume that the binary formation rate follows the cosmic star formation rate (SFR)~\cite{Vangioni:2014axa}. We obtain the merger rate by convolving the SFR with a standard $p(t_d)\propto 1/t_d$ time-delay distribution, assuming different minimum time delays for BBHs and BNSs~\cite{Dominik:2013tma, LIGOScientific:2016fpe, LIGOScientific:2017zlf,Meacher:2015iua, LIGOScientific:2017zlf}. Since massive black holes are expected to originate mostly from low-metallicity stars~\cite{LIGOScientific:2016fpe, LIGOScientific:2016vpg}, we further assume a metallicity cutoff in the BBH redshift distribution. We consider a $\log_{10}$-normal distribution for the stellar metallicities with standard deviation of 0.5 and a redshift-dependent mean value based on Ref.~\cite{Madau:2014bja}, but increased by a factor of 3 to account for local observations~\cite{Vangioni:2014axa, Belczynski:2016obo}. We then reweigh the merger rate for BBHs with at least one component of mass larger than $30\, M_\odot$ by the fraction of stars with metallicity $Z < Z_\odot/2$~\cite{LIGOScientific:2017zlf}. 

Current estimates of the CBC SGWB are affected by the large astrophysical uncertainties on merger rates. We set the overall normalization of the merger rate to match the local rates from the most recent LVK catalog~\cite{LIGOScientific:2021psn,LIGOScientific:2020kqk}. We characterize the uncertainty by choosing a fiducial value for the local BBH (or BNS) merger rate, and then allowing it to vary within the $90\%$ confidence interval inferred from the latest GWTC-3 catalog~\cite{LIGOScientific:2021psn}. For BBHs, we adopt a fiducial value of $28.3~\rm Gpc^{-3}yr^{-1}$, which corresponds to the latest best estimate for the \texttt{POWER+PEAK} model~\cite{LIGOScientific:2021psn}, and a confidence range of $[17.9,44]\, \rm Gpc^{-3}yr^{-1}$, which spans different models compatible with observations~\cite{LIGOScientific:2021psn}. For BNSs, since we use a more sophisticated mass distribution than the ones assumed in Ref.~\cite{LIGOScientific:2021psn}, we use the local merger rate estimate of $320~\rm Gpc^{-3}yr^{-1}$ from the previous GWTC-2 catalog~\cite{LIGOScientific:2020kqk} to allow for a more immediate comparison with recent forecasts for XG observatories~\cite{Borhanian:2022czq,Ronchini:2022gwk,Iacovelli:2022bbs}, and the $90\%$ confidence interval $[10,1700]\,\rm Gpc^{-3}yr^{-1}$ inferred from the GWTC-3 catalog~\cite{LIGOScientific:2021psn}.

We set the coalescence time $t_c$ of all GW signals to zero. We assume that all angles (coalescence phase $\phi_c$, right ascension $\alpha$, declination $\delta$, polarization angle $\psi$, and inclination $\iota$) are distributed isotropically and that all binary components are nonspinning, as we expect the effect of spins to be subdominant.

\begin{figure*}
\includegraphics[width=\columnwidth]{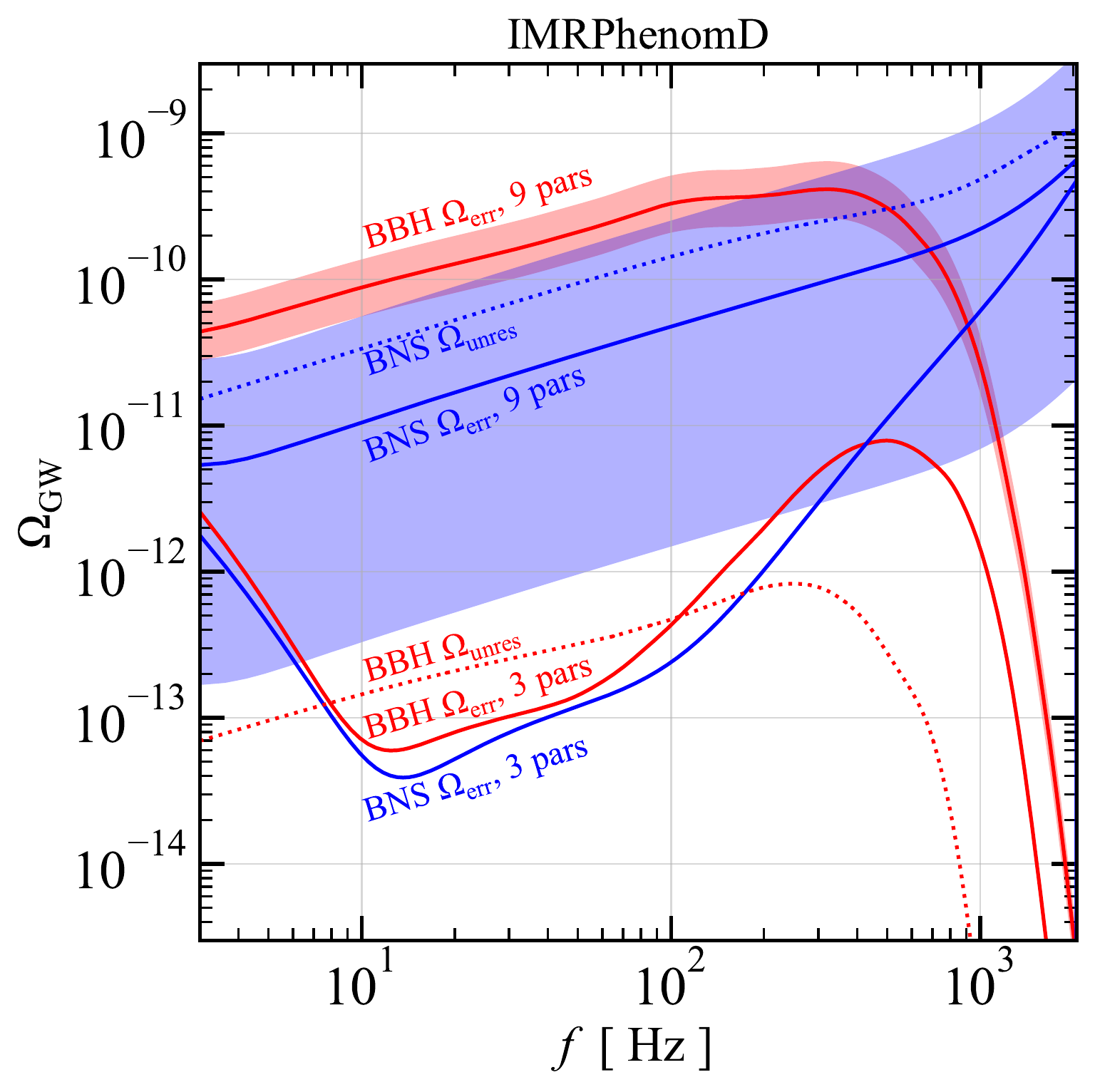}
\includegraphics[width=\columnwidth]{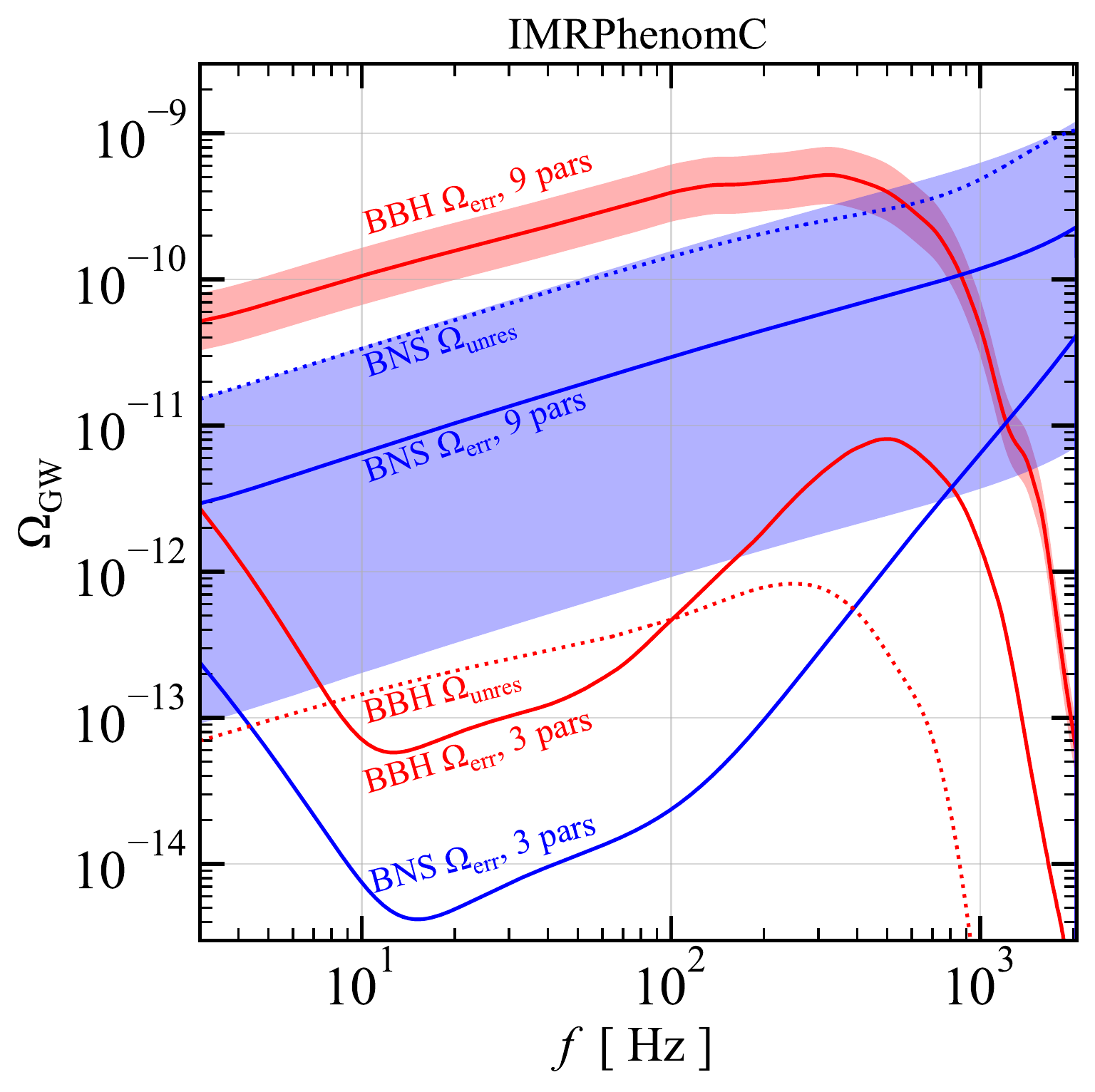}
\caption{
Unresolved background $\Omega_{\rm unres}$ (dotted lines) and error contributions to the background $\Omega_{\rm err}$ (solid lines) computed using both our 9-parameter recovery (``9 pars'') and the 3-parameter recovery (including only $\mathcal{M}_z$, $t_c$, and $\phi_c$) considered in previous work~\cite{Sachdev:2020bkk} (``3 pars'').  
All backgrounds are computed at the frequency-independent optimal $\rm SNR_{thr}$ for the 9-parameter case: 10 (for BBHs) and 20 (for BNSs) in the left panel; 12 (for BBHs) and 42 (for BNSs) in the right panel.
The shaded band around the 9-parameter $\Omega_{\rm err}$  shows astrophysical uncertainties on the rates. The worsening in $\Omega_{\rm err}$ due to including 9 parameters instead of 3 is quite dramatic and it is larger than astrophysical uncertainties, especially for BBHs.} 
\label{fig_OmegaErr_3vs9}
\end{figure*}

\noindent {\bf \em GW parameter estimation.} 
A population of $N$ unresolved CBC signals $\{h^i\}_{i=1}^N$ observed over a total time $T$ produces an energy density flux~\cite{Regimbau:2016ike,Sachdev:2020bkk}
\begin{equation}
F_{\rm tot}(f)=T^{-1} \frac{\pi c^{3}}{2 G} f^{2} \sum_{i=1}^{N}\left[|\tilde{h}_{+}^i(f)|^{2}+|\tilde{h}_{\times}^i(f)|^{2}\right] \,,
\label{eq_flux}
\end{equation}
where $f$ is the GW frequency, and $\tilde{h}_{+,\times}^i(f)$ are the two waveform polarization modes in the Fourier domain. This flux generates an energy-density spectrum~\cite{Allen:1997ad,Maggiore:1999vm}
\begin{equation}
\Omega_{\rm GW}(f) \equiv
\frac{f}{\rho_c} \frac{d\rho_{\rm GW}}{d f}(f) 
=  
\frac{1}{\rho_c c} f F(f) 
\,,
\label{eq_omegadef}
\end{equation}
where $\rho_{\rm GW}$ is the GW energy density and $\rho_c=3H_0^2/8\pi G$ is the critical density of the Universe. 
In general, the total CBC spectrum includes both detectable and undetectable sources. A standard technique to reduce the CBC foreground consists in fitting for the individually resolvable signals and subtracting them from the total~\cite{Regimbau:2016ike,Sachdev:2020bkk}. However, the recovery of detected signals is never perfect, due both to instrumental noise and systematics in waveform modeling~\cite{Cutler:2007mi}. The imperfect subtraction procedure leaves a residual for each removed signal, and the pile-up of these residuals generates an effective flux~\cite{Sachdev:2020bkk}
\beq
\begin{aligned}
F_{\rm error}(f) &= T^{-1} \frac{\pi c^{3}}{2 G} f^2 \sum_{i=1}^{N_{\rm res}} \left[ \left| \tilde{h}_+(\vec{\theta}^i_{\rm tr};f) - \tilde{h}_+(\vec{\theta}^i_{\rm rec};f) \right|^2 \right.  \\ 
&+ \left. \left| \tilde{h}_\times(\vec{\theta}^i_{\rm tr};f) - \tilde{h}_\times(\vec{\theta}^i_{\rm rec};f) \right|^2 \right] \,,
\label{eq_Ferr}
\end{aligned}
\eeq
where $N_{\rm res}$ is the number of resolved sources, $\vec{\theta}^i_{\rm tr}$ denotes the true parameters of each source, and $\vec{\theta}^i_{\rm rec}$ denotes the recovered parameters. Therefore, the total CBC background after subtraction is given by the sum of the astrophysical background from undetected signals $\Omega_{\rm unres}$ and the effective residual background $\Omega_{\rm err}$ obtained by substituting Eq.~\eqref{eq_Ferr} into Eq.~\eqref{eq_omegadef}.

We consider a \emph{fiducial} 3-detector network consisting of a $40$-km scale CE in Idaho (US), a $20$-km CE South in Australia, and an ET located in Italy~\cite{Borhanian:2020ypi}. For comparison, we also consider an \emph{optimistic} 5-detector network with four $40$-km CEs located at the LIGO Hanford, Livingston, India and KAGRA sites, plus one ET at the location of Virgo.
We include Earth-rotation effects for the longer BNS signals, and neglect them for BBHs.
We assume a GW signal to be detected if its network signal-to-noise ratio (SNR)~\cite{Schutz:2011tw,Maggiore:2007ulw} is above a certain threshold $\rm SNR_{thr}$. The undetected signals contribute directly to $\Omega_{\rm unres}$. For the detected signals, we estimate $\Omega_{\rm err}$ using the linear signal approximation~\cite{Finn:1992wt}. We assume the posterior probability distribution for each source to be a multivariate Gaussian centered at the true parameters $\vec{\theta}^i_{\rm tr}$ with covariance matrix $\Sigma = \Gamma^{-1}$, where $\Gamma$ is the network information matrix.
We then draw the recovered parameters $\vec{\theta}^i_{\rm rec}$ from each Gaussian posterior. 

Previous work~\cite{Sachdev:2020bkk} considered only $3$ parameters ($\mathcal{M}_z$, $t_c$, and $\phi_c$) in the calculation of $\Omega_{\rm err}$. Here we consider a larger set of $9$ parameters for the calculation of the information matrix for each CBC event:
\begin{equation}
\vec{\theta} = 
\left\{ 
\ln(\frac{\mathcal{M}_z}{M_\odot}), 
\eta, 
\ln(\frac{D_L}{\text{Mpc}}), 
\cos\iota, 
\cos\delta,
\alpha, 
\psi,
\phi_{c}, 
t_{c} 
\right\} \,,
\label{eq_info_paras}
\end{equation}
where $\mathcal{M}_z = \mathcal{M} (1+z)$ is the detector-frame chirp mass and $\eta$ is the symmetric mass ratio. To get a rough estimate of the effect of waveform systematics, we repeat the calculation with two different waveform models: {\tt IMRPhenomD}~\cite{Husa:2015iqa,Khan:2015jqa} as a ``fiducial'' reference model, and {\tt IMRPhenomC}~\cite{Santamaria:2010yb} for comparison. We compute the gravitational waveforms, the SNRs and the information matrices with the public package {\tt GWBENCH}~\cite{Borhanian:2020ypi}.

Increasing the value of $\rm SNR_{thr}$ increases the number of unresolved events (and thus $\Omega_{\rm unres}$), but decreases $\Omega_{\rm err}$, as the remaining detected signals have higher SNR and are better recovered. Hence, for each population one can determine an \emph{optimal} $\rm SNR_{thr}$ that minimizes the CBC background $\Omega_{\rm unres}+\Omega_{\rm err}$ left after the removal of resolved sources. We find that, to a very good approximation, the background is minimized for the same $\rm SNR_{thr}$ at all frequencies: for our fiducial detector network, this optimal $\rm SNR_{thr}$ is about 10 (20) for BBHs (BNSs) when we consider the {\tt IMRPhenomD} model, and 12 (42) for BBHs (BNSs) when we consider {\tt IMRPhenomC}. From now on we will show the estimated CBC SGWBs for these values of $\rm SNR_{thr}$, i.e., the lowest CBC backgrounds achievable with the fitting-subtraction procedure.

Figure~\ref{fig_OmegaErr_3vs9} shows the estimated $\Omega_{\rm unres}$ and $\Omega_{\rm err}$ for BBHs and BNSs computed at the optimal $\rm SNR_{thr}$ with the two waveform models. The difference between the residual backgrounds $\Omega_{\rm err}$ computed using our full 9-parameter recovery and those computed using the 3-parameter recovery of Ref.~\cite{Sachdev:2020bkk} is quite striking. The addition of amplitude parameters proves to be crucial, and $\Omega_{\rm err}$ becomes larger by several orders of magnitude, for both BBHs and BNSs. We find that the dominant contribution to $\Omega_{\rm err}$ arises from the coalescence phase $\phi_c$, as the error on this parameter becomes much larger once the correlations with amplitude parameters (particularly the polarization angle $\psi$) are taken into account. Another significant contribution arises from the luminosity distance $D_L$, which is known to be poorly constrained for a significant fraction of both populations, even with a network of XG observatories~\cite{Borhanian:2022czq,Ronchini:2022gwk,Iacovelli:2022bbs}.

The shaded bands in Fig.~\ref{fig_OmegaErr_3vs9} show the impact of astrophysical uncertainties on the local merger rates (note that this is a lower bound on astrophysical uncertainties, because the redshift evolution of the rates is even more poorly constrained). It is difficult to formulate reliable predictions for the CBC SGWB, especially for BNSs, where the background can vary by about two orders of magnitude. However, the increase in $\Omega_{\rm err}$ due to the addition of the amplitude parameters is even larger than the variability of the background due to the uncertain merger rates.
There is a clearly visible difference in the residual background $\Omega_{\rm err}$ predicted by the two waveform models, which is larger for BNSs than for BBHs. This difference, while small compared to current astrophysical uncertainties, highlights the importance of waveform systematics in data analysis and it will play a more prominent role in SGWB forecasts in the coming years, as new detections will steadily reduce the uncertainties in merger rates.

\noindent {\bf \em Background subtraction estimates.} 
\begin{figure}[t!]
\includegraphics[width=\columnwidth]{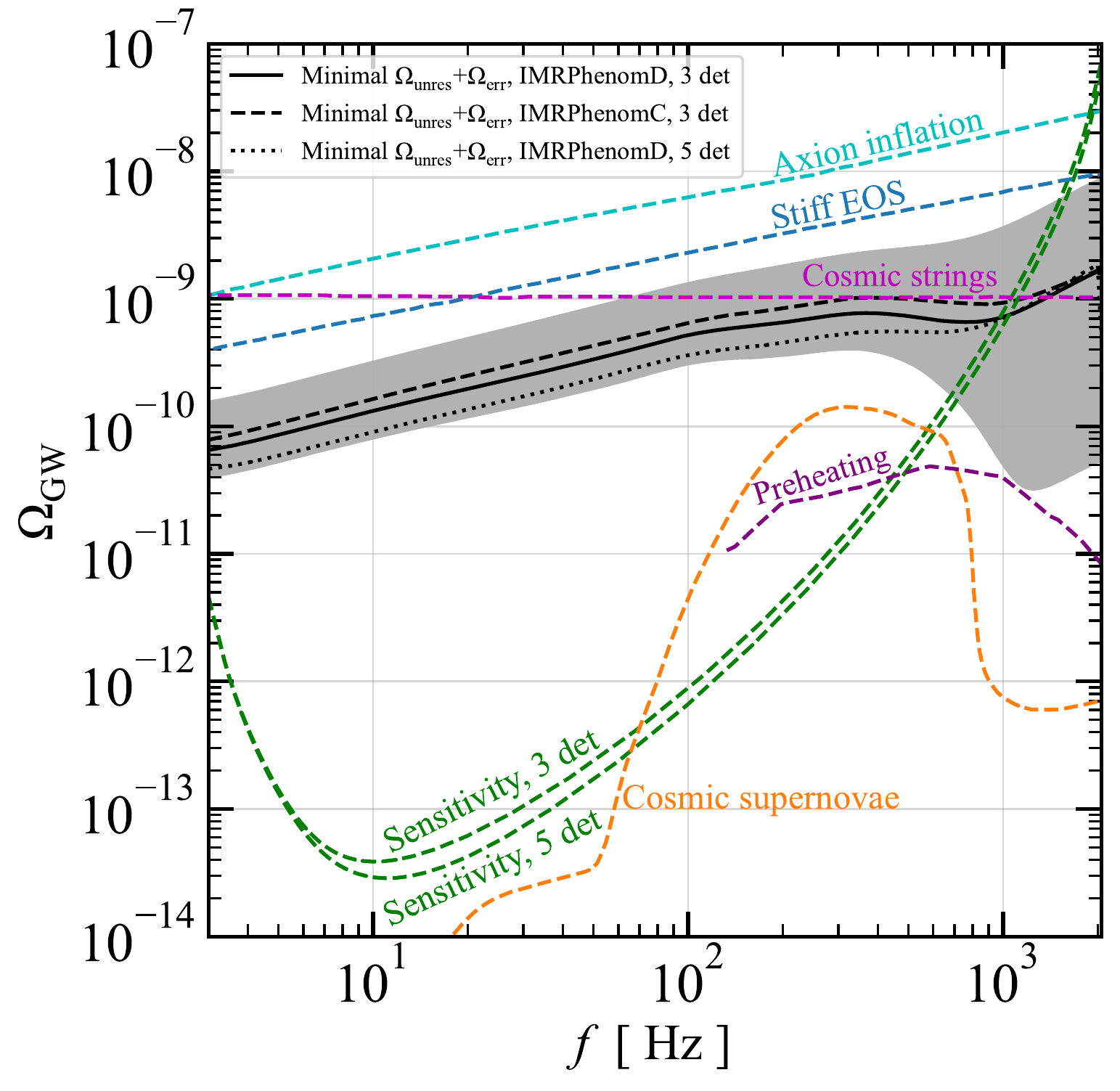}
\caption{
Minimal value of $ \Omega_{\rm unres} + \Omega_{\rm err} $ for BBHs and BNSs combined (black lines+grey band). This quantity is a good estimate of the total foreground for observing the SGWB produced by other possible sources, as labeled. For our fiducial 3-detector network, we show results for both {\tt IMRPhenomD} (solid) and {\tt IMRPhenomC} (dashed). We also show {\tt IMRPhenomD} results for the optimistic 5-detector network (dotted).
The astrophysical uncertainty band at 90\% confidence level is shown only in our fiducial case of {\tt IMRPhenomD} and 3 detectors, for clarity.
Green dashed lines are the sensitivities of XG detectors to SGWBs, assuming 1-year integration, in the absence of foregrounds from CBCs and other sources.
The SGWB from standard inflation~\cite{Grishchuk:1974ny,Starobinsky:1979ty,Grishchuk:1993te} ($\Omega_{\rm GW} \sim 10^{-15}$) is below the range of the plot. }
\label{fig_Omegas_final}
\end{figure}
Figure~\ref{fig_Omegas_final} shows the sum of the (minimized) $\Omega_{\rm unres}+\Omega_{\rm err}$ contributions from BBHs and BNSs for different waveform models and detector networks. The grey band shows the 90\% confidence level due to astrophysical uncertainties in our fiducial case ({\tt IMRPhenomD} and a 3-detector network).
We overplot the sensitivities of 3-detector and 5-detector networks to SGWBs, computed in the absence of foregrounds from CBCs and other sources, assuming the data is integrated for one year.  We also show the energy densities of SGWBs from other possible sources~\cite{Sachdev:2020bkk}, including: (i) axion inflation~\cite{Barnaby:2011qe}, (ii) post-inflation oscillations of a fluid with an equation of state stiffer than radiation~\cite{Boyle:2007zx}, (iii) a network of cosmic strings~\cite{Damour:2004kw, Siemens:2006yp, Olmez:2010bi, Regimbau:2011bm}, (iv) the most optimistic prediction from cosmic supernovae throughout the Universe~\cite{Finkel:2021zgf}, and (v) post-inflation preheating models aided by parametric resonance~\cite{Khlebnikov:1997di, Tilley:2000jh, Dufaux:2010cf, Figueroa:2017vfa}.  Note that the energy densities of these sources are model-dependent, and they could be larger or smaller depending on the choice of model parameters.

\noindent {\bf \em Conclusions and future directions.}
The main conclusion to be drawn from this work is that subtracting the SGWB foreground from BBHs and BNSs is much harder than previously estimated.
For our fiducial case of a 3-detector network and {\tt IMRPhenomD}, the BBH background subtraction only reduces it by a factor of 2--3. Crucially, the remaining BBH  background still overwhelms the BNS background at frequencies below hundreds of Hz. Similarly, the BNS background subtraction only reduces it by a factor of $\lesssim 2$.
As shown in Fig.~\ref{fig_Omegas_final}, even in the optimistic case of a 5-detector network the subtraction for the total (BBH+BNS) background only improves by a factor of $<2$ compared to our fiducial 3-detector network.
This can make searches for other sources of astrophysical and cosmological SGWBs very challenging.

Our rather pessimistic predictions for $\Omega_{\rm err}$ may be overcome by using other techniques to remove the foreground from resolved sources.
The residuals due to imperfect removal could be reduced by subtracting the component tangent to the signal manifold at the point of best fit. This approach has been first proposed by Ref.~\cite{Sharma:2020btq}, although more detailed investigations are needed to understand the extent of this reduction on realistic astrophysical catalogs.
Other possibilities include using Bayesian techniques to estimate the foreground and background signal parameters simultaneously~\cite{Biscoveanu:2020gds}, or exploiting the design topology of ET to construct a null stream~\cite{Freise:2008dk} that will help in understanding the foreground of CBC events~\cite{Regimbau:2012ir}.
One could also take advantage of the temporal and positional information of each event for a more precise subtraction.  
We hope that this study will motivate further work to assess the impact of these (and other) data analysis strategies on the detectability of SGWBs.

\noindent {\bf \em Acknowledgements.} 
%
We thank Sylvia Biscoveanu, Ssohrab Borhanian, Roberto Cotesta, Mark Hannam, and Alan Weinstein for helpful discussions. 
M.K. and B.Z. were supported by NSF Grant No.\ 2112699 and the Simons Foundation.
E.B., M.\c{C}. and L.R. are supported by NSF Grants No. AST-2006538, PHY-2207502, PHY-090003 and PHY20043, and NASA Grants No. 19-ATP19-0051, 20-LPS20- 0011 and 21-ATP21-0010. 
M.\c{C}.\ is also supported by Johns Hopkins University through the Rowland Research Fellowship. 
B.S.S. is supported by NSF Grants No. AST-2006384, PHY-2012083 and PHY-2207638. 
Part of E.B.'s and B.S.S.'s work was performed at the Aspen Center for Physics, which is supported by National Science Foundation grant PHY-1607611. This research was supported in part by the National Science Foundation under Grant No. NSF PHY-1748958.
This research project was conducted using computational resources at the Maryland Advanced Research Computing Center (MARCC).
This work was carried out at the Advanced Research Computing at Hopkins (ARCH) core facility (\url{rockfish.jhu.edu}), which is supported by the NSF Grant No. OAC-1920103.
The authors acknowledge the Texas Advanced Computing Center (TACC) at The University of Texas at Austin for providing {HPC, visualization, database, or grid} resources that have contributed to the research results reported within this paper \cite{10.1145/3311790.3396656}. URL: \url{http://www.tacc.utexas.edu}.


\bibliography{references}


\end{document}